# Non-Uniform Suction Control of Flow around a Circular Cylinder

James Ramsay[1], Mathieu Sellier[1, a)], Wei Hua Ho[2]

[1]Department of Mechanical Engineering, University of Canterbury, Christchurch, New Zealand
[2]Department of Mechanical and Industrial Engineering, University of South Africa, Pretoria, South Africa

a) Corresponding author: mathieu.sellier@canterbury.ac.nz

**Abstract.** In the present study, numerical investigations were performed with optimisation to determine efficient non-uniform suction profiles to control the flow around a circular cylinder in the range of Reynolds numbers $4 < Re < 188.5$. Several objectives were explored, namely the minimisation of: the separation angle, total drag, and the pressure drag. This was in an effort to determine the relationships between the characteristics of the uncontrolled flow and the parameters of optimised suction control. A variety of non-uniform suction configurations were implemented and compared to the benchmark performance of uniform suction. It was determined that the best non-uniform suction profiles consisted of a distribution with a single locus and compact support. The centre of suction on the cylinder surface for the optimised control, and the quantity of suction necessary to achieve each objective, varied substantially with Reynolds number and also with the separation angle of the uncontrolled flows. These followed predictable relationships, however. Non-uniform suction profiles were much more efficient at eliminating boundary layer separation, requiring the removal of less than half the volume of fluid as uniform control to achieve the same objective. Regardless of the method of control, less net suction was needed to minimise total drag than to eliminate separation. Though boundary layer separation could always be prevented, its elimination did not always result in an improvement in total drag, and in some circumstances increased it. An analysis of the drag components showed that, although the pressure drag could be substantially reduced by boundary layer suction, the disappearance of the separated region downstream resulted in faster flow over the cylinder and consequently a higher skin friction drag. The results show that the balance of drag components must be an important consideration when designing flow control systems and that, when done appropriately, substantial improvement can be seen in the flow characteristics.

## INTRODUCTION

The flow around a circular cylinder has been the subject of much interest for over a century and – for almost as long – so has the problem of controlling it. This bluff body flow can produce high drag and lift characteristics as well as a variety of undesirable flow phenomena – particularly vortex shedding. The behaviour of the flow and its impact on the aerodynamic characteristics of the cylinder vary widely with relatively small changes in Reynolds number ($Re = \frac{\rho U D}{\mu}$), complicating the search for optimal control parameters. Although much focus has been paid to minimising the drag or eliminating vortex shedding by flow control, the phenomenon of boundary layer separation and its characteristic parameter, the separation angle, has not attracted as much scrutiny. Despite the sustained interest in this field, the problem of optimal control for this flow is still very much open.

The flow around a circular cylinder offers a complex problem for the study of flow control because its features can be changed substantially by just varying the Reynolds number. In the range of $Re$ from $10^0$ to $10^7$, Williamson[1] distinguishes the resulting flows into nine regimes with unique features and phenomena. Zdravkovich[2] divides them more finely, with four additional regimes. At low $Re$, the flow is unseparated and steady ($Re < 6$), but in the range of $6 < Re < 47$ the boundary layer separates and a pair of vortices form behind the cylinder; the separation point moves toward the front of the cylinder with increasing $Re$. At about $Re = 47$, the well-known phenomenon of vortex shedding begins and continues in a two-dimensional manner up until $Re = 188.5$ where the vortex shedding begins to exhibit 3D features. With further increasing $Re$ the wake becomes even more complex as different areas of the flow transition to turbulence, beginning with the wake, moving up the shear layers, and finally the flow transitioning in the boundary layer, which marks the onset of the well-known 'drag crisis'. At this point, the separation point is delayed significantly, jumping much further aft and providing a substantial decrease in pressure drag. Clearly, any attempt to provide optimal flow control over this entire range would be exceedingly difficult if not impossible. However, by investigating small regions of the $Re$ range, relationships between the flow and control parameters can be established which may shed light on other regimes (or to entirely different body-fluid flows).

One of the important features of bluff body flows is the separation angle, $\theta_s$, which marks the point where the boundary layer separates from the surface as measured from the leading or trailing edge (LE or TE) of the body. It is commonly held that once the boundary layer has separated and become sufficiently established, the separation angle does not change significantly until the drag crisis at $Re \approx 2 \times 10^5$ where it jumps from about 100° to 60° (as measured from the TE, as in the rest of this paper) [3,4]. However, experiments over a wide $Re$ range by Weidman[5], and separately by Achenbach[6], show that the separation angle is a more active feature than commonly held, perpetually moving across the entire $Re$ range, and not monotonically. In many papers exploring the flow around circular cylinders, the separation angle is only described as a variable of secondary importance – usually in discussion of the features of the pressure profiles. However, as the separation angle is so intrinsically tied to the major phenomena of the flow (the wake, vorticity development, instability, etc.), it may be beneficial to control this feature directly. The pressure drag, $C_{d_p}$, which makes up almost all of the total drag on the cylinder at high $Re$ is predominantly caused by the separation of the boundary layer.

In 2004 Wu et al.[7] performed a review of experiments which had measured the separation angle of the cylinder in the range $0 < Re < 280$, and found that the results between studies deviated by as much as 10°. By their own experimentation and numerical simulation, the researchers found accurate values and determined the causes of the deviation – mostly due to different measurement techniques. This difficulty in measuring the separation angle, and the variety in measurements arising from its natural movement, contribute to why it is not always investigated. One method commonly used is to approximate that the separation point corresponds to the inflection point of the pressure curve, as employed by Fransson et al[8], but this may not be entirely accurate. However, the improvement of experimental techniques and equipment has now eliminated many of these problems, and these difficulties are not present in computational fluid dynamics (CFD). Therefore, there is no reason that the separation angle should not be a feature of particular interest.

To date, many active and passive methods of control have been investigated to control the flow around the circular cylinder. These range from simple geometric features such as splitter plates[9] and helical strakes[10] to complex active measures like plasma actuators[11] or magnetic fields[12]. Many of these are described in the Annual Review by Choi, Jeon and Kim[13] or more lately in the 2016 review by Rashidi et al.[14] Each of these methods have achieved some success at reducing drag or weakening vortex shedding, though often at significant cost. One of the simplest forms of flow control is boundary layer suction. This method removes the low momentum fluid particles at the surface, thus entraining higher momentum particles from the free-stream to reinvigorate the boundary layer, delaying separation. This method of flow control is as old as the boundary layer concept itself – with Prandtl testing his theory by experimenting on slot suction of a cylinder[15]. Nevertheless, it is still not a settled matter how this method can optimally control the flow around a circular cylinder, i.e. achieve the control objective with the least suction/fluid removal. Boundary layer suction has many advantages compared to other control methods. For one, the geometry of the body does not have to be changed (although the materials of the surface may). Further, the method is simple and practical – its parameters can be adjusted easily and have a wide range; there is no multi-physicality to this control. And of particular importance to this study, the method is well researched both experimentally and numerically.

Experiments on suction control began in the early 20th Century, and much interest was paid to this subject for the improvement of aerodynamic characteristics for aircraft during the Second World War[16]. Over this time, two main applications of suction control were investigated: uniform suction over the entire surface of a cylinder with porous walls – as researched experimentally by Thwaites and his colleagues[17,18] – and slot suction, where only part of the boundary layer is removed through a slot or series of slots in the cylinder surface. Pankhurst and Thwaites[17] found that sufficient uniform suction resulted in the flow approaching that seen for potential flow, with improvements in drag and elimination of vortex shedding. More recently, Fransson et al.[8] determined a relationship between the controlled flow using uniform suction on the cylinder and the uncontrolled flow at a different Reynolds number, and that these could be linked via the Strouhal number. However, the effective Reynolds relationship found by Fransson et al. is only applicable if the control does not entirely suppress vortex shedding which is a common objective for bluff body flow control. Uniform suction has its disadvantages, namely the inefficiency of removing material at all areas of the cylinder – even where it may not be necessary – and its limited control parameters. Slot suction is similarly disadvantaged, being limited in its location of application, the distribution profile of the suction, and the discontinuity of its nature.

A better approach to suction control combines the benefits of each of these methods – non-uniform suction. This method is applied similarly to uniform suction by use of a porous surface, however the suction is applied unevenly, with the possibility of any potential distribution over the surface – continuous or otherwise. Theoretically, this allows much more precise control of the flow, with the possibility to concentrate the suction control at critical areas of the surface and apply no control where it is unnecessary. Because there are so many potential profiles for this method, determining its "optimum" is not straightforward – even when only considering steady, time-independent control. Furthermore the influence of suction on the flow is nuanced. For example boundary layer suction can reduce drag in turbulent flows by relaminarising the flow thus decreasing the skin friction on the surface, whereas when applied to laminar flow it can have the opposite effect – increasing skin friction while decreasing the pressure drag.

Some papers on the subject of non-uniform suction have been published, particularly with the focus of utilising it in conjunction with feedback from the flow in order to mitigate the wide-range of potential implementations. Min and Choi[19] developed and employed sub-optimal feedback control with non-uniform suction and blowing (the optimisation is sub-optimal as it is over a finite, short time-frame). The study investigated the optimised flow and control parameters for three objectives: minimising pressure drag, minimising the difference of the surface pressure profile to that for inviscid flow, and maximising the square of the pressure gradient. The researchers successfully reduced the drag on the cylinder, and their results showed that the choice of objective had a large impact. The pressure drag objective did not result in the best drag reduction nor, surprisingly, did it result in the smallest pressure drag for a given set of conditions. The researchers showed that this was due to the balance of skin friction drag and pressure drag which were influenced in different ways by the control.

Though the researchers presented the separation points of the controlled flows in some instances, they did not investigate its potential as a control objective or how it relates to the optimised control parameters. They also limited their study to two Reynolds numbers, thus the potential changes of suction/blowing profiles change depending on the characteristics of the flow were limited. The sub-optimal suction/blowing profiles that achieved the most drag-reduction consisted of strong blowing near the rear of the cylinder and lesser suction near the top and bottom of the cylinder (90° and 270°). A later paper by Kim & Choi also found that at $Re = 100$, the flow was most sensitive to control by spanwise distributed slot suction[20]. In the remainder of that paper the control was applied at this 90° location for a wide range of Reynolds numbers, although the best location for suction control may move with $Re$.

Li et al. performed a similar numerical study as Min & Choi, and carried out a complete adjoint optimisation procedure with unsteady, time-dependent simulations. The researchers achieved a complete control of vortex shedding for up to $Re = 110$. They also found that the optimal controls were insensitive to initial conditions if the control was applied for time-scales longer than the vortex shedding period. The researchers used objectives for the error between the flow field and potential flow field, the enstrophy of the flow (to suppress vortex shedding), and the minimisation of drag. The separation angle of the flow was not investigated. Very recently, the optimum spanwise-varying suction/blowing control of a 3D circular cylinder in 2D flow was determined using eigenmode analysis by Boujo et al.[21] Due to the nature of this method, though, it can only be used to optimise for stabilisation or frequency modification which does not necessarily coincide with minimised drag or the elimination of separation. All of these studies were performed by numerical methods; non-uniform suction has not been explored substantially by physical experimentation.

These studies show that non-uniform suction can be used to efficiently achieve a variety of objectives in controlling the flow around circular cylinders. Nevertheless the relationships between parameters of the uncontrolled flow (particularly the separation angle) and the control to optimally minimise each objective are still not clear. In many cases systems for the modelling of sub-optimal or optimal control by suction/blowing were presented which can be replicated to achieve good results. However, an investigation of how the characteristics of the uncontrolled flow affect the optimised control generally, and whether there are any reliable relationships between the two, remains to be presented in a compact way. Furthermore, the focus has usually been on controlling the vortex shedding behind the cylinder, or minimising drag. The impact of the separation angle has gone relatively under-reported, and the question of whether it may be useful as a parameter or objective for direct control is also open. A general investigation of a variety of finite control configurations – how they are influenced by and themselves influence the flow – will provide useful information for those designing efficient flow control.

Therefore, the objective of the present study is to investigate a variety of suction control methods for a circular cylinder by numerical investigation, and to determine relationships between the optimised controls and characteristics of the uncontrolled flow. Particular attention is given to the separation angle as an output parameter and control objective. Several optimisation studies with a variety of control constraints and objectives were performed in the Reynolds range of $4 < Re < 188.5$ in order to best determine the relationships between the three factors of control – uncontrolled flow, control parameters, and the resulting flow. The $Re$ range employed offers a variety of flow regimes (unseparated flow, separated steady flow, and vortex shedding) while offering a simple, two-dimensional and fully laminar arrangement to carry out many simulations. Of particular interest were the impacts of different control configurations and objectives on the features of separation angle, drag components, pressure profiles, and the general features of the flow.

## METHODS AND MODELLING

### Geometry, mesh and solver methods

The governing equations for unsteady incompressible and isothermal viscous flow are described by the following:

$$\frac{\partial u_i}{\partial t} + u_j \frac{\partial u_i}{\partial x_j} = -\frac{\partial p}{\partial x_i} + \frac{1}{Re}\frac{\partial^2 u_i}{\partial x_j \partial x_j}, \tag{1}$$

$$\frac{\partial u_i}{\partial x_i} = 0, \tag{2}$$

$$\frac{\partial u_i}{\partial t} + u_j \frac{\partial u_i}{\partial x_j} = -\frac{\partial p}{\partial x_i} + \frac{1}{Re}\frac{\partial^2 u_i}{\partial x_j \partial x_j}, \frac{\partial u_i}{\partial x_i} = 0, \text{where}$$

$$\frac{\partial u_i}{\partial x_i} = 0, \text{where}$$

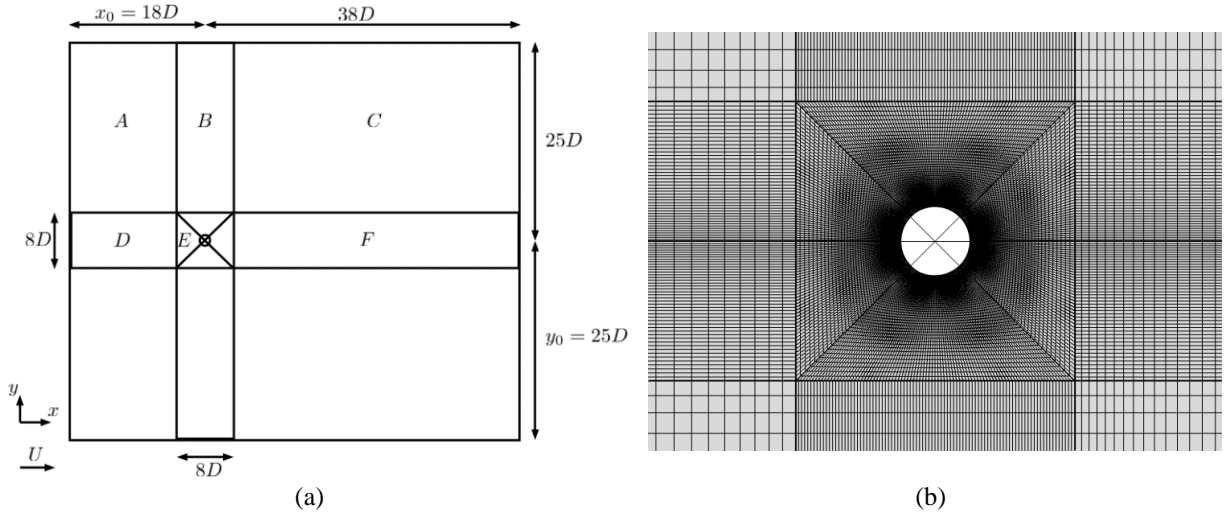

Figure 1: Sketch of (a) computational domain and (b) close-up view of element mesh around the cylinder

The inlet (left-boundary) was assigned a uniform flow boundary condition with velocity, $u = U = \frac{Re\,v}{D}, v = 0$. The upper and lower boundaries were modelled as no-slip moving walls with the same velocity profile as the inlet. This minimises any potential blockage effects that may result from the artificially bounded domain. The actual domain has blockage ratio, $BR = \frac{1}{50} = 0.02$ which was shown to reduce the error in separation angle to below 0.2% by Wu et al. A pressure outlet condition was imposed on the outlet (right-boundary) with zero relative pressure. The cylinder walls were modelled as a fixed-velocity outlet with defined normal outflow velocity, $u_n = v_w, u_t = 0$, where $u_n$ and $u_t$ are the normal and tangential velocity components at the wall, respectively. This boundary condition made it possible to define any suction or blowing profile on the cylinder wall by only changing the function that defines $v_w$, the suction velocity. In keeping with the terminology typically used in the literature, the non-dimensional suction coefficient, $c_q = \frac{v_w}{U} \times 100$, was used as the control parameter, from which $v_w$ was defined. In this paper $c_q$ with a lower case 'c' refers to the local suction coefficient at any particular point on the cylinder, while $C_q$ refers to the net suction coefficient of the cylinder as a whole, $C_q = \frac{1}{2\pi} \oint c_q d\theta$. The two definitions will be useful given that non-homogeneous suction profiles are the subject of this investigation. With this boundary condition, locations where no suction is applied have the same definition as a no-slip wall.

Table I: Characteristics of mesh found to be independent for steady-state and transient solutions.

| Region | Number of Grid Points (x × y) | Number of Elements | Average Quality | Minimum Quality |
|---|---|---|---|---|
| A | 20x20 | 400 | 1.000 | 1.000 |
| B | 90x20 | 1800 | 1.000 | 1.000 |
| C | 60x20 | 1200 | 1.000 | 1.000 |
| D | 20x90 | 1800 | 1.000 | 1.000 |
| E | 49x90 | 4410 | 0.843 | 0.500 |
| F | 60x90 | 5400 | 1.000 | 1.000 |
| Total | | 31640 | 0.904 | 0.500 |

## Suction control configurations

One of the objectives of this study was to investigate how non-uniform suction profiles might more efficiently control the flow around the circular cylinder than uniform suction. To explore this, three methods for applying non-uniform suction profiles were devised. These are summarised in Figure 2. The values underneath each configuration show the number of control parameters for each distribution.

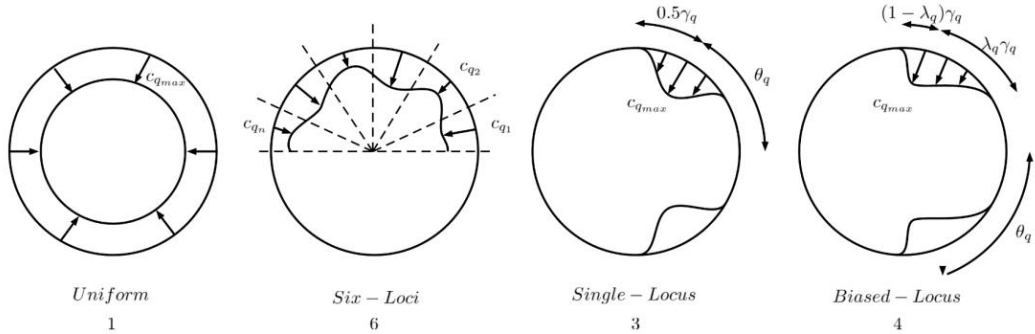

Figure 2: Schematic of the types of control methods investigated.

The "six-loci" control configuration can model non-uniform suction with a single locus (location of peak suction) or multiple, and thus has the most flexibility of the non-uniform suction profiles presented here. For this distribution, the upper half of the cylinder was divided into six equal segments and the suction coefficient defining the suction velocity at the centre of each was used as the control variables, $c_{q_n}$. As we were not concerned with the lift or lateral forces of the cylinder, the lower half of the cylinder control were set to mirror the upper half. To create a continuous suction profile from the six discrete values, pointwise interpolation with a continuous second-derivative constraint was applied. In this way, a suction profile with multiple loci of suction could be modelled. Earlier trials of using more "segments" found that six segments resulted in the best control. Control with 9 and 18 loci did not improve the control while taking longer to converge to an optimised solution. At $Re = 80$ for instance, these took 35% and 135% longer to solve than the six-loci system respectively, and achieved the objective with 6% more suction than the 6-loci solution.

Earlier investigations also found that the final suction profiles from the optimisation typically have only one locus of suction. The "single-locus" control was defined to generate this type of distribution using fewer parameters. Here three control parameters are used: the maximum local suction coefficient, $c_{q_{max}}$, centre of suction, $\theta_q$, and spread of suction, $\gamma_q$. To create a smooth suction profile, a cubic polynomial was defined from these parameters and the condition of zero-gradient was applied at the edges and centre of the profile. Thus, a distribution with compact support can be generated and its location can be varied easily by the optimisation system.

The "biased-locus" distribution was identical to the single-locus profile except for the addition of a fourth control parameter: a bias factor, $\lambda_q$. This allowed the suction profile to be asymmetrical. This does introduce the risk of a profile that is so steep it creates separation bubbles in the flow. In addition to these non-uniform suction profiles, uniform suction was also investigated. Here, only one parameter was necessary, $c_q$, which was held constant at all locations on the cylinder surface. In all cases, this parameter was limited such that the suction velocity at any point on the cylinder surface could be no greater than the magnitude of the free-stream velocity. This was in order to keep the problem bounded and also feasible for real implementation.

## Optimisation methods

Three different cost functionals were defined for the present study:

$$J_1 = \theta_s = \theta\left(\tau_w = \left.\frac{\partial u}{\partial r}\right|_{wall}\right) \tag{3}$$

$$J_2 = C_{d_t} = C_{d_p} + C_{d_f} = \frac{2}{\rho U^2 D}\oint\left(-p(\theta)|_{r=R} + \mu\left.\frac{-\partial u_t(\theta)}{\partial r}\right|_{r=R}\right)\cos(\theta)\,R\,d\theta \tag{4}$$

$$J_3 = \left|C_{d_p}\right| = \left|\frac{2}{\rho U^2 D}\oint -p(\theta)|_{r=R}\cos(\theta)\,R\,d\theta\right| \tag{5}$$

$J_1$ is the angle of separation relative to the trailing edge (TE). For this study, the separation point was located by determining the furthest upwind point where the wall shear stress is zero, $\max\left(\left.\frac{\partial u}{\partial r}\right|_{wall} = 0\right)$. This was checked by also evaluating where the flow reverses direction very near to the wall along a curve with a diameter of $1.002D$. Control that prevents separation will minimise $J_1$, reducing asymptotically to zero as the separation point reaches the trailing edge.

$J_2$ is the total drag coefficient, $C_{d_t}$, which is the sum of the pressure drag, $C_{d_p}$, and skin friction drag, $C_{d_f}$, coefficients. The total drag of the uncontrolled cylinder varies non-linearly in relation to $Re$ so we would not expect the prevention of separation to necessarily correspond to a minimisation of drag due to the several impacting factors contributing to this force. Because the suction control removes fluid and momentum from the domain, it is usually necessary to include a term for the "sink drag" that this generates. As the pressure and skin friction are integrated directly at the surface from the forces experienced by the cylinder, instead of using a momentum balance method, this sink drag is already incorporated into the $C_{d_p}$ and $C_{d_f}$ terms. Relating this to the drag measured by the momentum method gives the following equation, as seen in Beck et al.[22]

$$C_{d_t} = C_{d_p} + C_{d_f} = C_{d_w} + 2C_Q$$

where $C_{d_w}$ is the wake drag (as found by momentum theory) and $2C_Q$ is the value of the sink drag coefficient as derived, for example, in Tietjens & Prandtl[23].

$J_3$ is the pressure drag. It is typically assumed that the pressure drag arises solely from the separation of the boundary layer. The loss of momentum to the vortical structures that form as a consequence means that the pressure is not fully recovered on the leeward side of the body. However, pressure drag can arise even in an unseparated flow as there is momentum lost through the no-slip interaction with the wall. Because of this, the minimisation of pressure drag was investigated as its own objective to see how it compares to the separation objective results, $J_1$. Because, when situated on the front half of the cylinder, the suction can act similar to an inverted jet and thus provide a forward-biased pressure gradient (propulsion), the absolute value of the pressure drag is taken. The optimisation would otherwise try and maximise suction on the front-side to decrease pressure drag into substantial negative values. Taking the absolute value counteracts this so the optimisation determines the best control to reduce pressure drag to zero with the least suction. This is also why an objective for $C_{d_f}$ was not investigated. Any addition of suction at the boundary layer generates sink drag and increases skin friction, therefore the suction control that gives the best skin friction characteristics is no control.

In addition to these three, an additional objective was included in each of the studies. The best suction control was defined as that which would achieve the objective with least effort, therefore a secondary objective is necessary to measure the controller effort – the net suction. This objective is defined as the cost function, $J_w$, below. The overall cost functional for the studies investigated here, is therefore the sum of $J_w$ and one of the main objectives ($J_n$). As the efficiency of control is of secondary concern to achieving the actual flow characteristic objective, a scaling factor of 0.01 was employed in the addition of $J_w$ to the global objective as shown in Equation 5. It was found by trial that this was sufficient to be registered in the optimisation process, but only once the primary objective is achieved over the investigated $Re$ range.

$$J_w = C_q = \frac{1}{2\pi} \oint c_q(\theta) \, d\theta \tag{4}$$

$$J_{global} = J_n + 0.01 J_w \tag{5}$$

By experimenting with a variety of optimisation methods it was found that the gradient-free Nelder-Mead[24] (NM) method was effective and quick. The NM method is relatively insensitive to the parameter order and the typical optimisation process required about 100-300 iterations. Each iteration consists of solving the steady-state flow for the input parameters, evaluating the objective function, and adjusting the control parameters for the next iteration. Time for solving was dependent on the Reynolds number, taking longer for higher $Re$.

## RESULTS

### Validation of model

The model was validated by comparing the separation angles for the uncontrolled flows to those found in experiments and other numerical studies, as reviewed by Wu et al[7]. This is shown in Figure 3 (a). It can be seen in this figure that the time-dependent simulations model the flow accurately, with the time-averaged separation angles matching extremely well ($R^2 = 0.9993$). The instantaneous behaviour is also accurate, although the values from the literature for these values are not shown in the plot. In this study, the aerodynamic characteristics of the cylinder are of particular interest, therefore it was necessary to validate the measurements of the drag components also. Again, this was carried out by comparing the uncontrolled flow from time-dependent simulations with historical data[25–27]. These results are shown in Figure 3 (b) and show a good fit. The data from Wieselsberger[27] and Tritton[25] come from physical experiments, while the results from Henderson are from direct numerical simulations (DNS). The values from Henderson are used as the benchmark in this paper as the author provided good fits to his data. They are limited, however, in that Henderson only modelled the cylinder at $Re > 25$ so the fits may not be valid for the steady regimes.

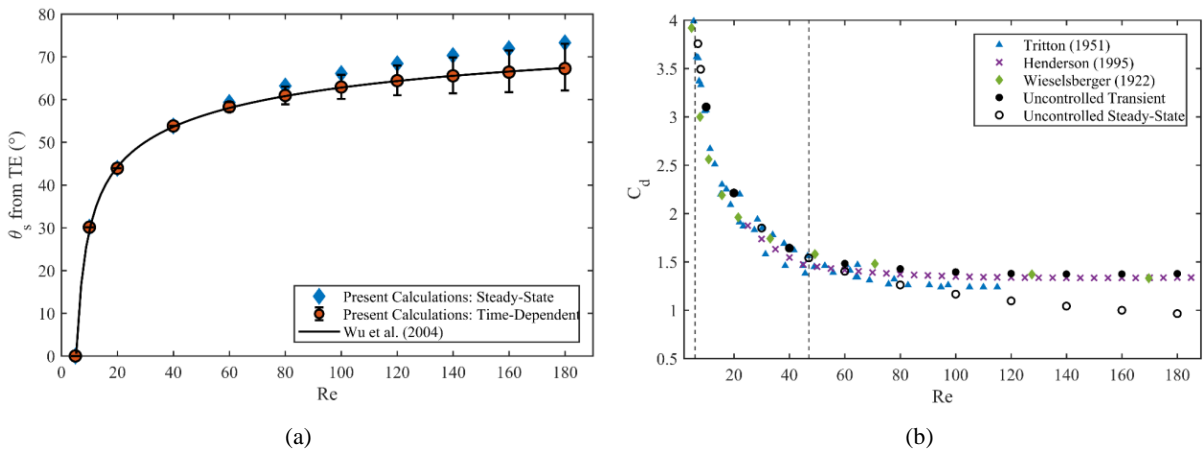

Figure 3: Validation of model by comparison to historical data for (a) the separation angle and (b) the time-averaged total drag coefficient. Here, the error-bars indicate the span of instantaneous values, while the points are time-averaged values.

In this investigation, we made the assumption that performing optimisation using the steady-state Navier-Stokes equations would result in control that, as it should stabilise the near-wake, is also effective in the true, unsteady flow. Therefore, it was necessary to determine how the uncontrolled flow is resolved using the steady-state equations, and how this compares to the literature. As can be seen in Figure 3 (a), the steady-state results diverge from the expected values within the vortex shedding regime ($Re > 47$) resulting in an $R^2$ value of 0.9073. This is unsurprising. However, these values do follow the same trend as the actual behaviour, and they result in earlier separation angles than reality. They provide a conservative approximation of the separation angle, therefore. Similarly, for the drag coefficient, the steady-state results diverge from the actual behaviour and underestimate the total drag.

It is known that sufficient boundary layer suction can stabilise an unsteady flow, particularly near the controlled surface. We make the assumption, therefore, that control that eliminates boundary layer separation will result in stable flows. On the other hand, it has been seen, such as in the work by Chomaz[28], that even steady parallel wakes are not necessarily stable and can become unsteady. It is shown later in this paper, that this assumption holds true for almost the entire investigated $Re$ range, and its effects are mostly felt in the far-wake than near the cylinder surface. The phenomena we are interested in, which arise at the cylinder surface and near wake, are therefore relatively unaffected. In fact, the maximum error in the optimised control flows was 3.29% error in the total drag at $Re = 180$, all other characteristic features had smaller differences between the steady and time-dependent models. Further details are given at the end of the Results section.

## Separation objective

In this section, we present results from the optimisation studies that employed objective to minimise the separation angle, $J_1$. We investigate the effectiveness of each suction control setup, with particular focus on the total suction coefficient, $C_q$, the centre of suction, $\theta_q$, the effect on drag and its components, and the pressure profile over the cylinder with and without control.

### A. *Comparison of control configurations*

For all control configurations, the objective of eliminating boundary layer separation was successfully achieved. A sample of the resulting flows can be seen in Figure 4 along with the instantaneous flow field of the uncontrolled case for comparison. It can be seen in these figures that the controlled flows all have a similar structure, with a much smaller, symmetrical wake. The streamlines illustrate how the freestream fluid is entrained as it passes the cylinder to replace the fluid removed through the suctioned surface. An important feature to note is that the velocity vectors of the flow near the top and bottom of the cylinder are much larger in the controlled cases. As there is no longer stagnated or separated flow downstream, the fluid can move more quickly over the cylinder – more like the potential flow.

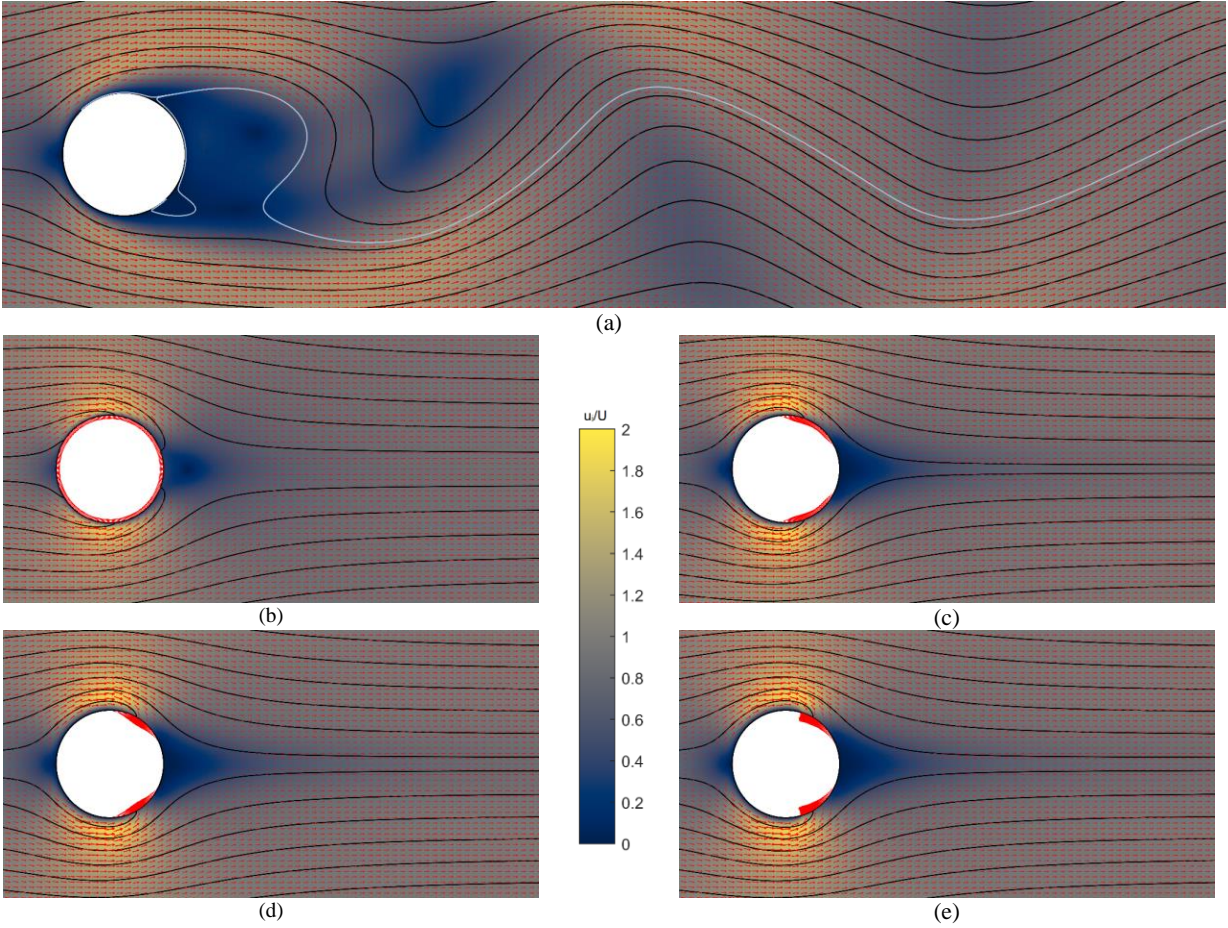

Figure 4: Instantaneous flow fields for controlling separation angle at $Re = 120$: (a) uncontrolled case, (b) uniform suction, (c) six-loci control, (d) single-locus, and (e) biased-locus distribution. The streamlines, non-dimensionalised velocity vectors ($\frac{u_i}{U}$) and non-dimensionalised velocity surfaces ($\frac{u_i}{U}$) are shown. The dense red arrows on the cylinders map the suction profiles.

The amount of suction required to eliminate separation is much greater for the uniform case than the non-uniform methods of control. The plots in Figure 5 show the suction quantity coefficient, $C_q$, against the Reynolds number and against the initial separation angle before control is applied, $\theta_{s_0}$. It is clear from this figure that uniform suction requires much more control effort to eliminate separation compared to any of the other methods. In all instances (except the trivial non-separated cases), the control effort is at least twice that of non-uniform suction.

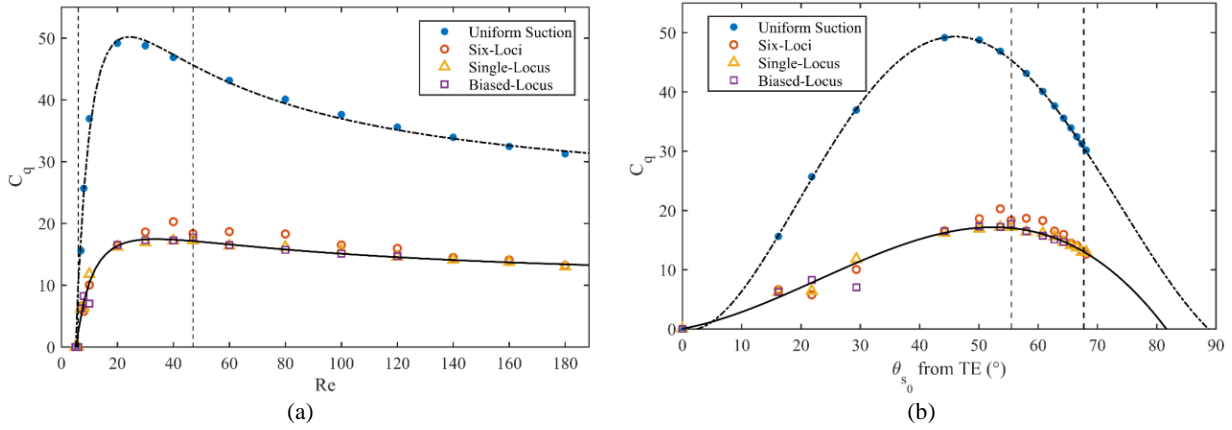

Figure 5: Global suction coefficient for optimised control to prevent separation plotted against (a) the Reynolds number, (b) the separation angle of the uncontrolled flow. The dashed vertical lines indicate regime changes of the uncontrolled flow.

These plots exhibit relationships between the uncontrolled flow features and the necessary control effort. For all control configurations, the amount of suction required to eliminate separation increases in the vortex-pair regime ($Re < 47$) up to a point, after which it decreases with increasing $Re$. Figure 5 (a) provides rational fits for the uniform and single-locus data each with a 2$^{nd}$ degree numerator and denominator. For the relationships with the initial separation angle in Figure 5 (b), polynomial fits were more appropriate. Extrapolating these trends to higher Reynolds numbers suggests that the suction control effort would plateau near $C_q = 23$ and $C_q = 10$ for uniform and single-locus suction respectively. Pankhurst & Thwaites[17] found that a suction quantity of $C_q\sqrt{Re} \geq \frac{30}{\pi}$, was required to eliminate separation on a cylinder fitted with a splitter plate in the $Re$ range of $10^4 - 10^5$. Extrapolating the present results to this regime, suggests a suction coefficient of only $C_q\sqrt{Re} \approx \frac{14}{\pi}$ is necessary, contrary to their experimental results[a]. Likewise, Pankhurst & Thwaites' relationship cannot be extended to the present regime, $Re < 188.5$, as it would suggest a suction coefficient of $C_q = 150.9$ would be required at $Re = 40$ to stabilise the flow, which is not the case. The Reynolds number does not contain sufficient information about the drastic changes in flow features to allow for these relationships to be extended.

On the other hand, the separation angle is a feature of the uncontrolled flow that itself is altered with the regime changes. Since, there is a trend between the uncontrolled separation angle, $\theta_{s_0}$, and the required control, it may be possible to extend the relationship with $\theta_{s_0}$ into higher $Re$ ranges. After all, the mechanism by which the suction reinvigorates the boundary layer should also remain the same whenever the boundary layer and the shear layers directly adjacent to it are still laminar, in other words, almost up to the transition to turbulence at $Re \approx 2 \times 10^5$.

It can be seen in Figure 5 that the best results are usually achieved with the biased-locus distribution. Though, the improvement in $C_q$ compared to the single-locus profile is marginal – less than 3% at all $Re > 10$, and in some instances the single-locus gave a better result ($Re = 80, 100$). Furthermore, the biased-locus distribution has potential issues with discontinuities in the model due to the steep suction profiles it can generate (as seen at $Re = 120$ in Figure 4). Finally, the optimisation process is most stable and quicker for controls with few parameters. Therefore, since there seems to be little advantage in using the more complex biased-locus distribution over its symmetrical variety, the rest of the results will be concerned with the single-locus profiles only.

### B. Optimised suction profiles

An interesting result from the optimisation studies for this objective was how much the optimised profiles move and morph depending on the Reynolds number. At low $Re$, the suction profiles are narrowly spread and positioned near the leading edge of the cylinder. As $Re$ increases, the profile spreads wider and moves further leeward on the cylinder. This shift can be seen in Figure 6 where a sample of the results at different $Re$ are shown for the single-locus control. In addition, lines marking the uncontrolled separation angle and the centre of suction are shown for each profile. Evidently, there is a relationship between the suction centre, $\theta_q$, and the Reynolds number (and by extension the uncontrolled separation angle as well). Figure 7 (a) and (b) plot these relationships respectively.

---

[a] Pankhurst &Thwaites defined $C_q$ as the flow rate through the porous wall divided by $(UD)$, i.e. $C_q = \frac{v_w}{U}\pi$ according to our notation, hence the introduction of the $\pi$ term to their equation in this text.

From Figure 7, it is clear that there is a strong relationship between $Re$, $\theta_{s_0}$ and $\theta_q$. Similar discussion can be made about these relations as for $C_q$ in the earlier section. These points will not be repeated beyond stating that the results confirm a dependence on the Reynolds number and also on the uncontrolled separation angle for the application of suction control; that it is not constant at all $Re$; and that it would be of interest to see how these relationships change when extended to higher $Re$. There appears to be less correlation with the spread of suction, $\gamma_q$, which has an average value of 75°.

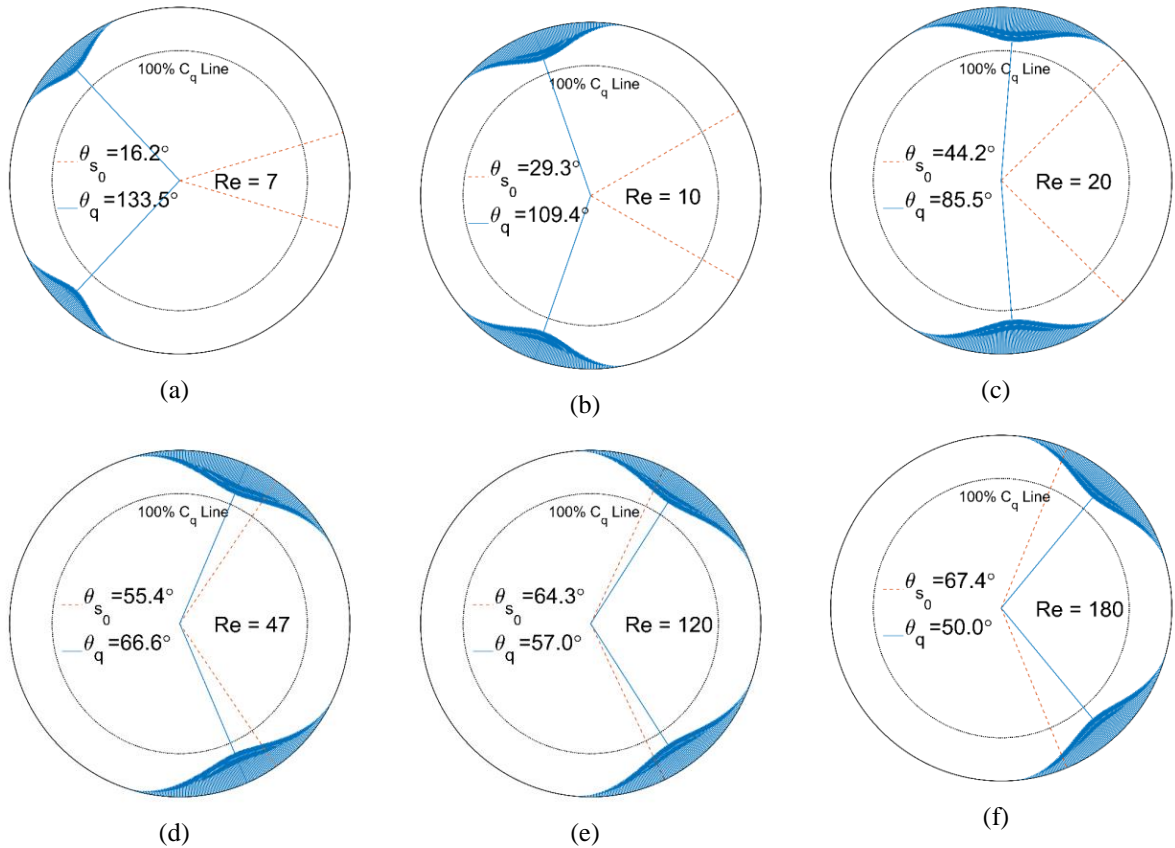

Figure 6: Variations in optimised suction profiles for single-locus control at various Re. The inner dotted circle marks where the local suction coefficient is 100, i.e. $v_w = U$. The uncontrolled (initial) separation angle, $\theta_{s_0}$, and resulting centre of suction, $\theta_q$, are also plotted and their values displayed.

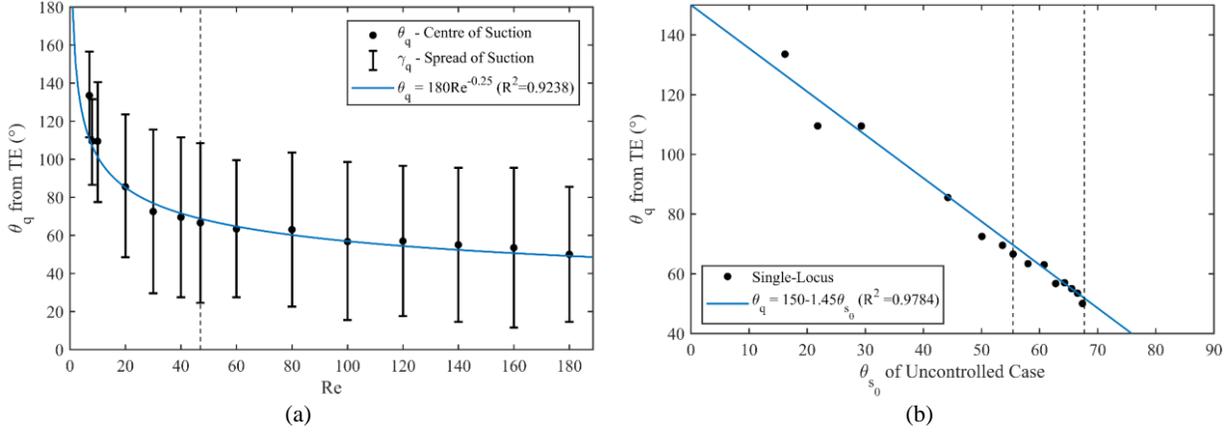

Figure 7: How the non-uniform suction profile with single-locus control moves with (a) Reynolds number and (b) initial separation angle. The vertical lines mark the regime changes of the uncontrolled flow.

## C. Effect on drag

The drag components for the final controlled flows were evaluated. These are plotted in Figure 8 alongside the values for the uncontrolled case using the relationships taken from the numerical analysis by Henderson[26]. There are several features to note here, in particular: the general trend of the total drag, $C_{d_t}$, and the behaviour of its two components, $C_{d_f}$ & $C_{d_p}$.

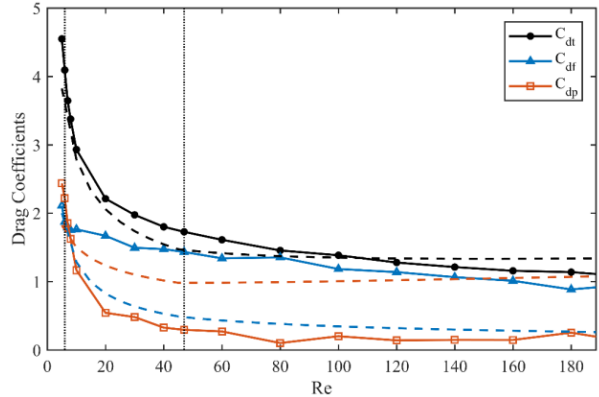

Figure 8: Drag components for the final controlled (solid lines) and uncontrolled (dashed lines) flows. The vertical dotted lines delineate the uncontrolled flow regime changes.

Figure 8 shows that all the values follow similar trends to the uncontrolled case: beginning very high and decreasing with a power-law relationship with increasing $Re$. However, where one might have expected the elimination of boundary layer separation to improve the drag, it is evident that this is not always the case. For all $Re < 100$, the drag is worse for the controlled case than the uncontrolled case; only above $Re = 100$ is improvement seen. The explanation for this behaviour can be found by analysing the components of the drag, shown in the same figure.

The first thing to note is the shift in pressure drag. Since the pressure drag is predominately contributed to by the loss of momentum to the boundary layer and the vortices that form in the separated region, one might have expected the pressure drag to be eliminated entirely, along with the boundary layer separation. This is clearly not the case. The optimisation studies searched for control parameters that eliminated separation with minimal suction effort, therefore the boundary layer is not entirely removed. With this objective, despite $\theta_s$ being successfully reduced to zero, momentum from upstream is still lost to the boundary layer, so the pressure is not fully recovered over the leeward side. This can be seen visually by the slow velocity region in the wakes of the controlled flows in Figure 4, and is also shown in the pressure profiles given in the following section. Nevertheless, the pressure drag is substantially reduced, particularly in the vortex shedding regime.

Counteracting the improvement of this one component, is a worsening of the other: the skin friction drag. With the removal of the separated region of flow, the boundary layer has a higher velocity across the entire surface of the cylinder, as was highlighted by the increased velocity vectors in Figure 4. This higher boundary layer velocity results in a stronger shear force and, correspondingly, a greater skin friction drag. At low $Re$, where the viscous effects of the flow are more important, this increase in skin friction drag can overwhelm the improvement in pressure drag. This imbalance results in a worsening of the total drag. In this case, for the objective of eliminating separation using single-locus control, this counterproductive imbalance occurs for all $Re < 100$. Only above $Re = 100$, where the inertial effects are sufficiently dominant and the improvement in pressure drag is more substantial, does this control work in favour of reducing total drag.

### D. Pressure profiles

To complete the analysis of the controlled flow behaviour and characteristics, the pressure coefficient profiles are provided in Figure 9. In the first plot, Figure 9 (a), the time-averaged pressure coefficients for the uncontrolled flow are compared to values from experiments in the literature. These values, labelled 'Zdravkovich', in the plot are taken from the curve fit by Zdravkovich[2] to experimental values from Thom[30], and Homann[31] in the $Re$ range $36 < Re < 107$. Hence, why at low Re values, $Re \leq 20$, the pressure coefficients are seen to differ quite substantially.

In Figure 9 (b), the controlled flow is compared to the uncontrolled flow for $Re > 40$. There are several features to highlight here. Firstly, the pressure profile of the controlled flow fills out more to become similar to the profile given by potential flow theory. The minimum pressure coefficient is much lower, though, with values between -3.75 and -4.5. This is similar to what was seen in the experiments by Pankhurst and Thwaites[17]. The lower pressure coefficient implies that the flow is being accelerated more than it would in the inviscid case. In addition, as the flow is no longer inhibited by the separated region, the pressure at the leading edge moves closer to $C_p = 1$. Finally, it is important to note that the plateau of $C_p$ near the trailing edge is not eradicated – momentum is still lost to the boundary layer, hence non-zero pressure drags were observed in Figure 8.

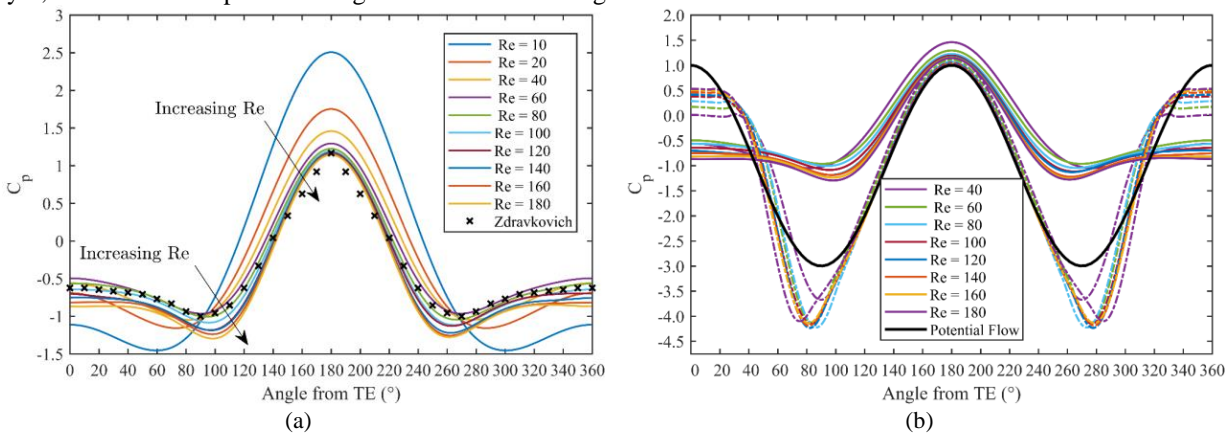

Figure 9: Pressure coefficient profiles over surface of cylinder for (a) uncontrolled case, and (b) both uncontrolled (solid line) and controlled (dot-dashed line) cases.

## $C_{d_t}$ and $C_{d_p}$ objectives

So far, we have discussed the results of control optimised for eliminating separation. As was described earlier, however, much of the present literature is more concerned with vortex shedding and drag coefficients than the boundary layer behaviour directly. Now, we consider the effects of changing the objective of optimisation to minimise the total drag or pressure drag using the single-locus control configuration.

### A. Comparison with $\theta_s$ objective

Figure 10 below shows the resulting flow fields at $Re = 120$ for the optimised control found for each objective using the single-locus method. It is clear from this figure that the control effort required to achieve each objective differs significantly, as does the behaviour of the resulting flow. While the uncontrolled flow field was taken from a time-dependent simulation, the other figures were taken from the final stage of the optimisation process and thus with a steady-state condition. Naturally, the steady-state flows are symmetrical, therefore, and there are no lateral movements in the wake as with the uncontrolled flow.

The first feature to note in Figure 10 is the difference in suction profiles for each control objective. As was shown earlier, the suction to eliminate separation at $Re = 120$ was spread wide and focussed near the trailing edge, with a relationship close to $\theta_q = 180 \, Re^{-0.25}$. The suction profiles for the drag objectives are narrower and closer to the top and bottom of the cylinder (90° and 270°). It can be seen visually, that the amount of suction, $C_q$, is much smaller for these objectives also – particularly for the total drag objective in Figure 10 (c). This makes sense given what the earlier results revealed about the balance of drag components: while boundary layer suction can reduce the pressure drag, it comes at the cost of increasing the skin friction drag. The separation angle objective often resulted in a net increase in drag because the suction was too strong, so it is appropriate that the suction profiles optimised to minimise total drag employ less suction.

The resulting flow fields for each objective differ significantly. The most obvious feature is the wake size – both its length and width. The wake in these figures can be considered the paler blue region centred on the trailing edge, bordered by the dark blue lines where the flow is stagnant. This delineates the two shear layers of reversed flow in the wake, and forward flow outside. With this definition, the separation objective flow in Figure 10 (b) has no real wake as it has no reversed flow, only stagnating fluid. On the other hand, the total drag objective has the longest wake. Both the $C_{d_p}$ and $C_{d_t}$ objective controls have a similar wake width, corresponding to a separation angle about 45° from the trailing edge.

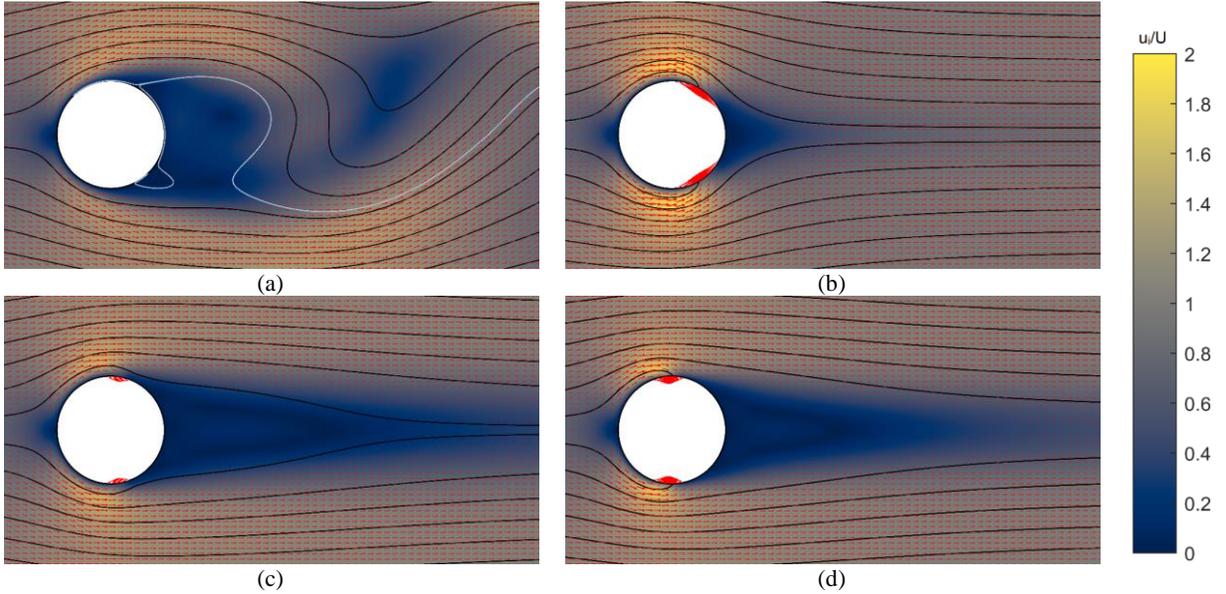

Figure 10: Instantaneous flow field at $Re = 120$ showing non-dimensionalised velocity vectors and surfaces ($\frac{u_i}{U}$), as well as streamlines from the inlet for the (a) uncontrolled case, and controlled for minimising (b) separation angle, (c) total drag and (d) pressure drag objectives.

Figure 11 demonstrates how the optimised control differs depending on the objective of optimisation. The amount of suction and the centre of suction are plotted against the Reynolds number, as in Figure 5 and Figure 7. As with $J_1$, clear trends can be seen in the optimised control parameters for the other objectives. Contrary to the trend seen for minimising $\theta_s$, however, Figure 11 (a) shows that the amount of suction required to achieve the drag objectives decreases with increasing $Re$ in all flow regimes. This figure also shows a levelling off at $C_q = 5$ for the amount of suction to minimise total drag within the vortex shedding regime.

Figure 11 (b) shows that the centre of suction, when optimised for the drag objectives, follows a similar trend to that for $J_1$. A power law is seen for each, and approximate fits are given in the legend of that figure. These fits have been rounded, so are not necessarily the best fits for the data, but help to make comparisons easier. It can be seen that the drag objectives result in suction profiles located closer to the leading edge, and begins to level off near 90°. This fits with what Kim and Choi[20] found for $Re = 100$, with the drag on a cylinder improving most by slot-suction and blowing when the slots were located between 80° and 100°.

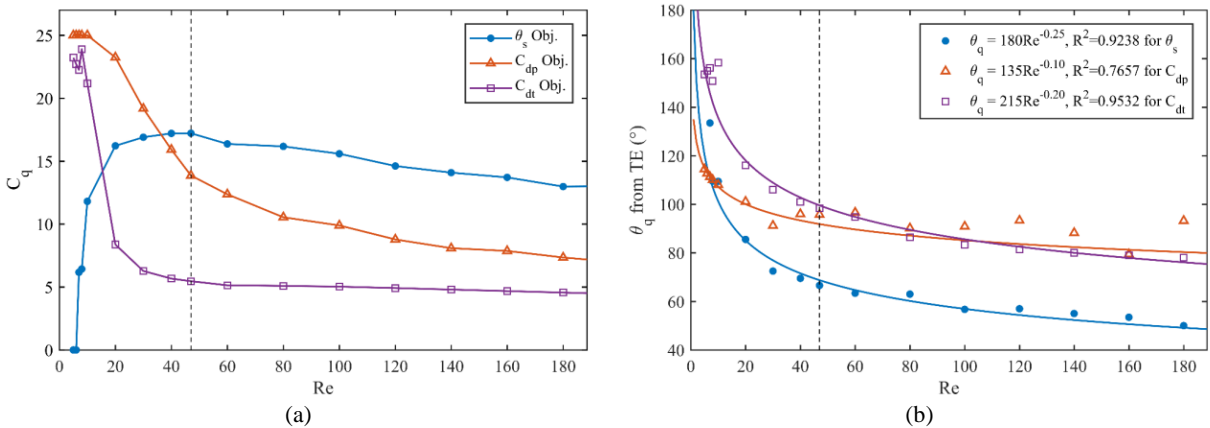

Figure 11: Effect of objective on optimised suction characteristics: (a) the amount of suction, and (b) the location of suction.

Since the drag objectives result in controlled flows that still have boundary layer separation, Figure 12 shows the separation angle for each controlled case in comparison to the time-averaged value for the uncontrolled case. Again, we see a levelling off of the data for the total drag objective. Here the separation angle plateaus near the 45° mark, thus the difference in separation angle before and after control is applied increases through the vortex shedding regime.

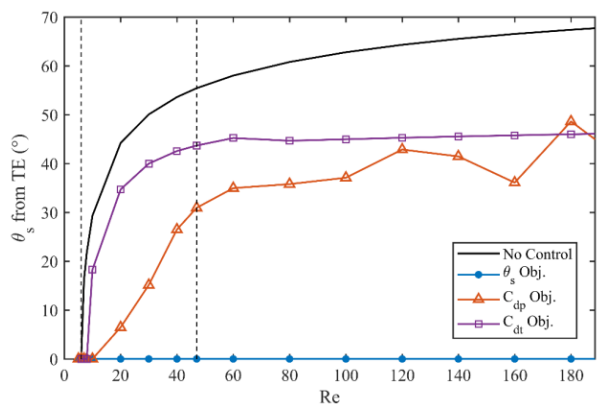

Figure 12: Separation angle for the resulting controlled flow for each objective. Uncontrolled values taken from Wu et al[7].

### B. Resulting drag characteristics

The effect of control on the components of drag is of great interest for these objectives, particularly how the skin friction and pressure drag changes are balanced to achieve the minimum total drag. The total drag of the flows with optimised single-locus control are displayed in Figure 13. Here, the relationships from Henderson[26] are also plotted for comparison to the uncontrolled case.

From this figure, it can be seen that the improvement to drag is much more substantial for these objectives than when eliminating separation – particularly in the vortex shedding regime. Before the von Karman street begins forming at $Re = 47$, there is little improvement in the total drag from either the $C_{d_t}$ or $C_{d_p}$ objectives. Although, whereas the separation objective in many instances resulted in a worsened total drag coefficient, the $C_{d_t}$ objective resulted in controlled flows that were never worse than the uncontrolled flow. On the surface this is unsurprising, as the optimisation algorithm would return zeroed control parameters if no suction configuration could improve upon the uncontrolled flow – thus the total drag should never be higher than the uncontrolled case for this objective – however, as can be seen in Figure 11 and Figure 12, significant control was applied in every instance. This shows that the drag can always be improved, or at least matched, by the application of non-uniform suction control in the entire investigated Reynolds range. It also suggests that very different control parameters can result in flows with near-identical macroscale features. The $C_{d_p}$ objective had slightly worse drag characteristics in the unseparated and vortex pair regimes, but large improvements in the vortex shedding regime. This fits with the observations from the separation objective results.

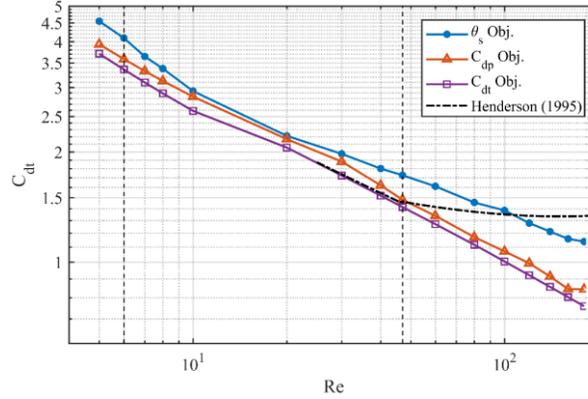

Figure 13: Total drag of the optimised, controlled cases compared with the time-averaged value for the uncontrolled flow.

The particularly important result shown in Figure 13 is the vast improvement in drag that occurs in the vortex shedding regime. The maximum decrease in drag was achieved at $Re = 180$ for the total drag objective, with a reduction of 0.578 (43.3%). This is in comparison to the modest 0.198 (14.8%) reduction seen for the separation angle objective for the same conditions. Here, while the uncontrolled drag curve begins to level off before increasing again, the drag on the optimised cylinder continues to decrease at the same rate as at lower $Re$. This means that, once vortex shedding has begun, as $Re$ continues to increase, the maximum improvement in drag also increases.

It is interesting that the characteristics for the total drag objective continue along the same trend as the 'subcritical' uncontrolled flow (subcritical here meaning before the onset of vortex shedding, $Re < 47$). The transition from steady separated flow to transient vortex shedding has a big impact on the aerodynamic characteristics of the uncontrolled cylinder: the pressure drag stops decreasing with $Re$ and begins to increase, while the skin friction drag coefficient continues to decrease at near its prior rate. The fact that the drag-optimised cylinder is unaffected by this transition, with the total drag continuing to decrease at its prior rate, suggests that the significant change imposed by the control is the counteracting of the pressure drag contribution attributable to the dynamic wake. It will be shown in the following sections that this improvement is not due to the use of steady-state equations to resolve these flows. Time-dependent simulations with the same control as used in the steady-state simulations result in nearly the same value (a maximum error of 3.29%).

## Verification by time-dependent studies

One of the potential limiting factors of the optimisation procedure in this study was the use of steady-state solvers for flows that, when not appropriately controlled, are unsteady ($Re > 47$). To test the validity of the assumptions in the optimisation process, time-dependent studies were performed with the optimised control found for each objective at $Re = 60$ and $Re = 180$. This tests the two extremes of the 2D vortex shedding regime. In each case, the vortex shedding was allowed to fully develop on the uncontrolled cylinder, before the control was ramped up linearly from 0% to 100% over the course of $t = 5T$, where $T$ is the period of vortex shedding given by Roshko's equation $St = \frac{D}{TU} = 0.212 - \frac{4.5}{Re}$. The controlled flow was then allowed to fully develop. The time was non-dimensionalised according to $t^* = t\frac{U}{x_0}$, where $x_0$ is the distance from inlet to the centre of the cylinder (see Figure 1).

Figure 14 shows the results of these verification studies, with the steady-state results shown as markers on the right vertical axis. As can be seen in Figure 14, the results fit well between the time-dependent and steady-state controlled results. In Figure 14 (b) it can be seen by the small oscillation of the separation angles on the upper and lower surfaces that the flow is not entirely steady at $Re = 180$ with the drag-optimised controls. For these controls, the wake became unsteady at some distance from the cylinder surface. Despite this, the major features of the cylinder flow were accurately evaluated by the steady-state studies, particularly the drag components and separation angles. The maximum error in these values was for the pressure drag objective at $Re = 180$. Here the maximum error was for the total drag, being lower by 0.106 (3.29%) in the steady-state simulations than in reality.

In addition to some of the transient details being missed in the steady-state simulations, another feature was absent for the flow at $Re = 60$ with the $C_{d_t}$ objective control applied. That feature is a small separation bubble which formed where the suction was applied in the time-dependent study. The impact of this can be seen in Figure 14 (a) where the separation angle was detected first at 93.39° where the separation bubble formed. This bubble was very small, as shown by the inset of this figure, and had no significant impact on the flow or the later major separation point (which is also plotted on this figure).

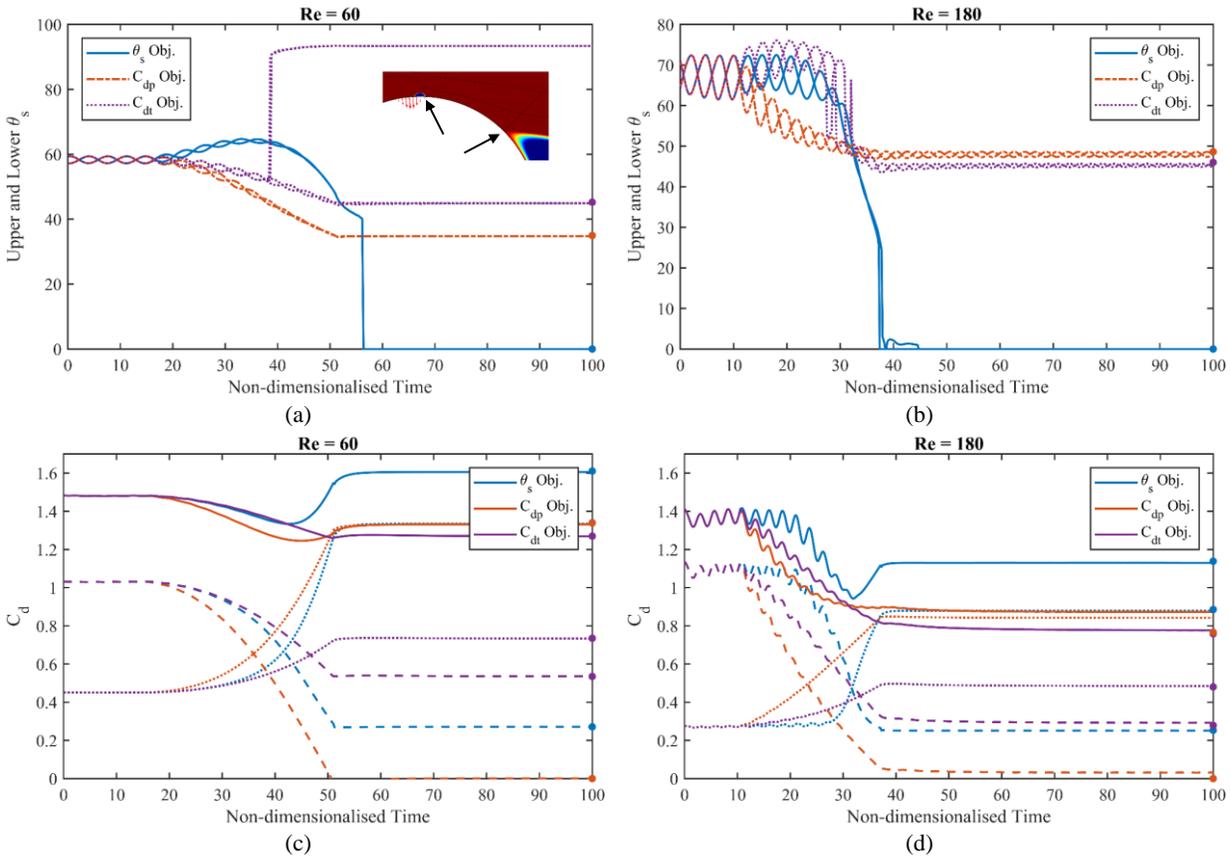

Figure 14: Results from time-dependent studies of the optimum single-locus control for each objective, showing (a-b) the instantaneous separation angles, and (c-d) the drag components. In (c-d) the dotted lines mark the skin friction drag, and the dashed lines mark the pressure drag. Inset on (a) is a tangential velocity surface for the $C_{d_t}$ objective at $Re = 60$ showing the separation bubble and main separation, that give rise to two sets of measurement for $\theta_s$.

The aim of quickly designing effective boundary layer control using optimisation with steady-state assumptions was achieved, and the results proved reliable. These time-dependent models took 11x and 43x longer to solve than the steady-state simulations at $Re = 60\ \&\ 180$ respectively. They also required a solution for the fully-developed uncontrolled flow for their initial conditions, and must be run until fully developed – which for less effective control may take even longer. Furthermore, each optimisation process took between 100-300 iterations, therefore substantial savings in time were made by applying the steady-state assumption for the optimisation. With this approach, the major characteristics of the flow were accurately modelled and the transient studies verified that the optimised control parameters found by steady-state simulations are effective for the real flows.

In addition to these time-dependent studies, the effectiveness of the optimisation was tested by performing a full parametric study for the single-locus configuration at $Re = 180$. With steps of 20° for the centre of suction and spread of suction, and steps of 10 for the suction strength, it was found that all objectives successfully converged to the appropriate minimum – the point where the primary objective was at its lowest in this parameter space with the least controller effort, $C_q$. These results are summarised in Table 2 below.

**Table 2:** Comparison of optimised results and parametric study global minima for Re=180

| Objective | Optimisation Result | | | | Parametric Minimum | | | |
|---|---|---|---|---|---|---|---|---|
| | $c_{q_{max}}$ | $\theta_q$ | $\gamma_q$ | $J_{global}$ | $c_{q_{max}}$ | $\theta_q$ | $\gamma_q$ | $J_{global}$ |
| $J_1$ – separation angle | 66.29 | 50.02 | 70.48 | **0.1298** | 65 | 60 | 100 | **0.1806** |
| $J_2$ – total drag | 40.98 | 78.03 | 40.04 | **0.8039** | 35 | 80 | 40 | **0.8081** |
| $J_3$ – pressure drag | 98.02 | 93.23 | 26.98 | **0.0734** | 65 | 80 | 40 | **0.0789** |

# CONCLUDING REMARKS

In this study, we have performed an optimisation of the control parameters for non-uniform boundary layer suction in an effort to achieve a variety of objectives for the flow around a circular cylinder, namely: eliminate separation, $\theta_s$, minimise total drag, $C_{d_t}$, and minimise pressure drag $C_{d_p}$. Numerical simulations were performed on the flow around a circular cylinder in the Reynolds range $4 < Re < 188.5$, and the Nelder-Mead method was used to determine the best parameters for the control. To thoroughly investigate potential implementations of non-uniform suction, several control models were tested. The "single-locus" control – a suction distribution with compact support and single locus – proved to be superior. It provided efficient, finite dimensional control and achieved the optimisation requirements. The results of this study provide a variety of insights into the problem of flow around a circular cylinder and flow control.

Firstly, clear relationships have been derived between characteristics of the uncontrolled flow and the control parameters for a variety of important objectives. The substantial improvement in efficiency of non-uniform suction in comparison to uniform suction suggest that this is the best form of boundary layer suction control. However, the dependency of its location on the parameters of the flow may make it impractical in many circumstances. Since $C_q$ and $\theta_q$ change with predictable trends though, such as $\theta_q = 180 \, Re^{-0.25}$ for the $\theta_s$ objective, this issue may be mitigated.

Secondly, the balance of drag components was shown to be a vital consideration in the suction-controlled flows. Eliminating separation did not always reduce the total drag as the skin friction component increases due to the fuller boundary layer profile and the incorporation of sink drag. Above $Re = 100$, however, the total drag of the controlled fully-attached flow is reduced as skin friction becomes less significant. It may be reasonably expected that for all $Re > 100$, the total drag can be reduced by effective non-uniform suction as the skin friction component of drag naturally becomes less consequential as viscous effects are proportionally reduced. The controlled flow to minimise total drag almost always required less suction than to eliminate separation, due to the balance of effects suction has on the pressure and skin friction components. The drag-optimised flows followed trends similar to that seen for a cylinder with steady flow in the unsteady regime (steady, unstable flows as described by Fornberg[32]).

Finally, the investigation into the importance of the separation angle, both as an uncontrolled parameter of the flow and as an objective of control, has yielded valuable results. It has been determined that the relationships between the flow and the optimised control could equally be expressed as a relationship with $\theta_{s_0}$ than with $Re$. In many cases this results in less convoluted connections. For example, the $\theta_q - \theta_{s_0}$ relationship was linear. Less dependence on the initial separation angle was seen for the drag objectives, however it plays a significant role as an outcome of the controlled flows, as seen by the repeated $\theta_s = 45°$ for the drag-optimised flows.

Overall, non-uniform suction profiles have been developed that effectively achieve a variety of objectives; strong relationships between features of the uncontrolled flow, the optimised control parameters, and the resulting flow have been reported; and finally, it has been shown that the separation angle is an important parameter for flow control – both as an input/ouput parameter, and potentially as an objective of the control.

# ACKNOWLEDGMENTS